\pdfoutput=1 
\documentclass[conference]{IEEEtran}
\IEEEoverridecommandlockouts

\usepackage[T1]{fontenc}
\usepackage[utf8]{inputenc}

\usepackage{amsmath,amssymb,amsfonts}
\usepackage{graphicx}
\usepackage{textcomp}
\usepackage{xcolor}
\def\BibTeX{{\rm B\kern-.05em{\sc i\kern-.025em b}\kern-.08em
    T\kern-.1667em\lower.7ex\hbox{E}\kern-.125emX}}

\usepackage{array}
\usepackage{makecell}    
\usepackage{xspace} 
\usepackage{booktabs} 
\usepackage{multirow}
\usepackage{adjustbox}
\usepackage{caption}
\usepackage{subcaption}
\usepackage[lined, noend, linesnumbered]{algorithm2e}
\usepackage{algcompatible}
\usepackage{algpseudocode}
\usepackage[most]{tcolorbox}
\newtheorem{definition}{Definition}

\usepackage{comment}
\usepackage{graphicx}

\usepackage{xcolor}

\newcommand{\ie}{i.e.}

\usepackage[colorlinks,
citecolor=violet,
urlcolor=violet,
linkcolor=violet,
bookmarks=false,
hypertexnames=true]{hyperref}

\usepackage{flushend}

\definecolor{darkgreen}{rgb}{0.01, 0.75, 0.24}
\definecolor{darkblue}{rgb}{0.01, 0.01, 0.54}
\definecolor{purple}{rgb}{1, 0, 1}

\newcommand{\tup}[1]{\left<{#1}\right>}

\newcommand{\checkTemp}{\mathsf{checkTemperature}}
\newcommand{\checkLasso}{\mathsf{checkLasso}}
\newcommand{\checkCycle}{\mathsf{checkCycle}}
\newcommand{\isHot}{\mathsf{isHot}}
\newcommand{\reportViol}{\mathsf{reportViolation}}

\usepackage[
  sorting=none,
  style=ieee, 
  backend=biber,
  isbn=false,
  url=false,
  doi=false,
  mincrossrefs=100,
  maxnames=12,
  giveninits
]{biblatex}

\DeclareUnicodeCharacter{0301}{*************************************}
    
\begin{document}

\title{Liveness Checking of the HotStuff Protocol Family}

\author{\IEEEauthorblockN{J\'er\'emie Decouchant, Burcu Kulahcioglu Ozkan, Yanzhuo Zhou \thanks{This document is a preprint of a paper that has been accepted for publication at the IEEE PRDC 2023 conference. An extended version of this work can be found in Yanzhuo Zhou's MSc thesis, which is available at \url{http://resolver.tudelft.nl/uuid:dd3c4287-ac18-4e9e-852d-24c3bf584dc7}.}}
\IEEEauthorblockA{\textit{Department of Software Technology} \\
\textit{Delft University of Technology}\\
Delft, the Netherlands \\
J.Decouchant@tudelft.nl, B.Ozkan@tudelft.nl, Y.Zhou-21@student.tudelft.nl}
}

\maketitle

\thispagestyle{plain}
\pagestyle{plain}

\begin{abstract}
Byzantine consensus protocols aim at maintaining safety guarantees under any network synchrony model and at providing liveness in partially or fully synchronous networks. 
However, several Byzantine consensus protocols have been shown to violate liveness properties under certain scenarios. Existing testing methods for checking the liveness of consensus protocols check for time-bounded liveness violations, which generate a large number of false positives.  
In this work, for the first time, we check the liveness of Byzantine consensus protocols using the temperature and lasso detection methods, which require the definition of ad-hoc system state abstractions. 
We focus on the HotStuff protocol family that has been recently developed for blockchain consensus. In this family, the HotStuff protocol is both safe and live under the partial synchrony assumption,  
while the 2-Phase Hotstuff and Sync HotStuff protocols are known to violate liveness in subtle fault scenarios. 
We implemented our liveness checking methods on top of the Twins automated unit test generator to test the HotStuff protocol family.
Our results indicate that our methods successfully detect all known liveness violations and produce fewer false positives than the traditional time-bounded liveness checks. 
\end{abstract}


\section{Introduction}
\label{sec:intro}

Byzantine fault tolerance (BFT) is a paradigm that gives distributed systems the ability to tolerate a limited proportion of arbitrary faults (\ie, Byzantine faults) such as equivocation (\ie, sending conflicting messages to different nodes) and loss of internal state. 
In particular, BFT consensus protocols aim at solving the consensus problem among $n$ nodes that might include up to $f$ faulty nodes. BFT consensus protocols aim at ensuring the safety and liveness properties. Safety ensures that the correct nodes always decide on the same value, while liveness ensures that the protocol always eventually progresses. 

Guaranteeing the liveness of a BFT consensus protocol is a difficult and error-prone process. 
For example, 2-phase HotStuff is a consensus protocol that Yin et al. discuss for pedagogical purposes and that could initially be considered live~\cite{hotstuff}. However, a particular scenario is shown to prevent the system from making progress, as nodes alternatively vote on two conflicting blocks.   
A similar attack called the force-locking attack~\cite{force_locking}, breaks both the safety and the liveness of a preliminary version of the Sync HotStuff protocol by maliciously delaying messages. 
These examples of liveness violations call for effective testing methods that will assist researchers and developers in detecting and tracing them.

Previous testing works on consensus have mostly focused on analyzing crash-tolerant protocols. 
For example, Jepsen testing tool~\cite{jepsen_verify} simulates network partitions for distributed systems, and it has detected several violations in the consensus systems~\cite{jepsen_cass, jepsen_etcd, jepsen_tw}. Twins~\cite{twins} is one of the few testing systems that have been specifically designed to test the safety of BFT consensus protocols under Byzantine scenarios. Twins can detect safety violations using scenarios that involve only a few communication rounds. However, most existing testing systems do not check for liveness violations, which require the generation of infinite executions.

A common approach to finding liveness violations is to check for \emph{bounded} liveness, i.e., checking whether the properties are satisfied within a bounded amount of time\cite{killian2007life, musuvathi2008fair}. To do so, the programmer sets some bounds for an event to happen and reports the executions that exceed the specified thresholds. In the case of consensus protocols, correct processes should accept the same value within a certain delay or within a given number of execution steps. However, it is difficult for developers to correctly estimate adequate bound values, in particular in real-world production-level consensus applications. Low bound values lead to false negatives while using very large bound values incurs high running times.  

Specific liveness testing methods have been proposed for distributed systems. 
\emph{Temperature-based} detection algorithms~\cite{killian2007life, partialcaching} maintain a temperature variable that is increased each time the system transitions to a hot state, whose definition is system-specific. 
The \emph{lasso detection} approach~\cite{partialcaching} relies on state caching to identify whether a system reaches the same hot state multiple times and therefore discover potential liveness violations. 
Little work has been done to effectively apply these techniques to test liveness violations in the blockchain consensus. 
In this work, we apply temperature and lasso detection methods to test the liveness of BFT consensus algorithms. We focus on the HotStuff family of protocols that have been designed for Blockchain consensus, which are sometimes called streamlined as they rely on a leader to reach a linear communication complexity and use a lock-step approach. We believe that our approach can be generalized to all protocols that rely on views, and in particular to all partially-synchronous protocols that rely on a leader~\cite{bftsmart,damysus,repucoin,pistis}.  

In summary, this work makes the following contributions:
\begin{itemize}
\item We define the notion of \emph{partial state} and \emph{hot state} for the HotStuff family of protocols, which are required for the implementation of the temperature and lasso detection methods. 
\item Then, we present a variant of the lasso detection approach, which, differently from previous works, does not use a controlled scheduling environment to check for the existence of lassos.
\item We also describe how to extend the Twins framework to support the temperature and lasso detection methods for the liveness testing of BFT consensus algorithms. 
\item We evaluate the performance of our testing methods on state-of-the-art protocols from the HotStuff protocol family and compare their accuracy to one of the classical bounded-liveness checking methods. 
\end{itemize}

This paper is organized as follows. 
Section~\ref{sec:related} discusses the related work. 
Section~\ref{sec:protocol} provides some necessary background on protocols from the HotStuff family and on liveness violations that have been described in the literature. 
Section~\ref{sec:method} overviews the temperature and lasso-detection based liveness checking methods. Section~\ref{sec:our-method} provides the definition of hot states for streamlined protocols and describes our extensions to the liveness checking methods for these protocols. 
Section~\ref{sec:impl} discusses the implementation of our testing methods on Twins, and Section~\ref{sec:eval} presents our empirical evaluation. 
Finally, Section~\ref{sec:conc} concludes the paper.   
\section{Related Work}
\label{sec:related}

In this section, we discuss previous works on detecting liveness violations in BFT consensus protocols and testing BFT consensus systems.

\emph{Liveness violations in BFT consensus protocols:}
Berger et al.~\cite{pbft_read_attack} recently demonstrated how optimization of read-only requests could lead to liveness violation in the seminal PBFT protocol~\cite{PBFT}. The scenario they discuss isolates a subset of clients forever and disturbs the execution of the voting and view-change protocols. 
It is also well-known that network delays can thwart the progress of partially synchronous algorithms. For example, a faulty leader in PBFT can withhold its \texttt{PRE-PREPARE} message until a timeout triggers a view-change, and the scheduler delays the receipt of \texttt{VIEW-CHANGE} message, leading to the new leader always having difficulty catching up with the latest progress~\cite{honey_badger}. 
The bouncing attack~\cite{bouncing_attack_casper} permanently damages the liveness of Casper FFG~\cite{casper} under the partially synchronous model. This attack is caused by malicious processes withholding or delaying a proportion of votes, leading to two conflicting chains alternately locking the current block. 
The preliminary version of Tendermint suffers from a similar violation in \cite{bc_in_wild} when two different processes alternatively lock their own proposed value at different heights. 
Amoussou-Guenou et al. further exhibit this vulnerability in~\cite{DBLP:conf/opodis/Amoussou-Guenou18} with a complicated 7-round scenario. We believe violations in Casper and Tendermint are variants of attacks in the HotStuff family protocols. In~\cite{revisit_zyzzyva}, a liveness attack on FaB is presented~\cite{fab}. This attack is simple: a faulty leader equivocates conflicting proposals and assembles a quorum certificate conflicting with that contained in the new-view messages. After a view change, the system is stuck because it can not vouch for the conflicting values. Bravo et al.~\cite{DBLP:journals/dc/BravoCG22} formally demonstrate that 2-Phase HotStuff can only ensure liveness using a timeout mechanism within a view. The leader should not make a new proposal until receiving the information from all correct processes. Otherwise, the liveness breaks as described in~\cite{hotstuff}.

\emph{Testing BFT consensus:}
There is a large body of works that propose new consensus algorithms, make them robust against Byzantine faults~\cite{DBLP:conf/nsdi/ClementWADM09}, or model-check the correctness of consensus algorithms \cite{tla, DBLP:conf/isola/0001W18, DBLP:journals/lmcs/0001LSW23}. In this section, we focus on the most related work on testing BFT consensus implementations.

Several existing methods for testing consensus systems focus on analyzing crash-fault tolerant protocols and exercise different executions of the systems under asynchrony, network faults, and crash process faults~\cite{url/jepsen, DBLP:conf/sigmod/AlvaroRH15, DBLP:journals/pacmpl/DragoiEOMN20}.   
Targeting BFT systems, \textit{BFTSim}~\cite{DBLP:conf/nsdi/SinghDMDR08} explores the system's behavior under unexpected network conditions and faults using a network simulator. 
\textit{Turret}~\cite{turret} detects performance attacks on BFT systems by generating Byzantine attack scenarios with malicious message deliveries, including message dropping, delay, duplication, and diversion. 
Several works~\cite{modelling_testing_NUS,zermia,DBLP:journals/corr/abs-2303-05893} provide testing frameworks that can model and inject network and Byzantine faults into the executions of BFT protocols.
\textit{Netrix}~\cite{DBLP:journals/corr/abs-2303-05893} provides a domain-specific language and a controlled networking environment that allows programmers to specify restrictions on the generated executions or implement their unit tests with network and Byzantine faults. 

\textit{Twins}~\cite{twins} systematically generates test case scenarios with Byzantine faults and explores them. It models Byzantine behaviors using twin copies of the processes, i.e., processes with the same identities and credentials. It runs the cluster with twin replicas and network partition, where the twin replicas exhibit Byzantine behaviors such as equivocation, double voting, and loss of internal state, which causes them to forget their voted values. 
Recent work \textit{ByzzFuzz}~\cite{byzzfuzz} generates test executions with randomly sampled network and Byzantine process faults. It models Byzantine faults using small-scope mutations to the original contents of the protocol messages and randomly injects a parameterized number of mutations. 
The goals of Twins and ByzzFuzz are orthogonal to our work: they provide test case generation methods rather than methods for checking liveness. 
Our liveness checking methods can be incorporated into any testing framework to check the liveness of the explored test executions.
In this work, we implement our methods on top of Twins, which already provide supports for testing the HotStuff protocol~\cite{url/hsrepo}.

\section{The HotStuff Protocol Family}
\label{sec:protocol}

In this section, we provide an overview of three streamlined blockchain consensus protocols: HotStuff, 2-Phase HotStuff, and Sync HotStuff. Whenever relevant we also discuss the liveness and safety issues that might affect them. 
These protocols present different combinations of liveness and safety properties, which are summarized in Table~\ref{tab:protocols}, and allow us to test our liveness checking methods. 

\begin{table}[b]
\normalsize
\centering
\renewcommand{\arraystretch}{1.2}
\begin{tabular}{ |c|c|c| } 
\hline
Protocol & Safety & Liveness \\
\hline
HotStuff~\cite{hotstuff} & Yes & Yes \\
2-Phase HotStuff~\cite{hotstuff} & Yes & No \\
Sync HotStuff~\cite{sync_hotstuff} (early version) & No & No \\
\hline
\end{tabular}
\vspace{1mm}
\caption{Summary of the safety and liveness guarantees of the three protocols from the HotStuff family that we consider.}
\label{tab:protocols}
\end{table}

\subsection{HotStuff}

\begin{figure}[t]
  \centering
    \includegraphics[bb=0 0 416 220,width=\linewidth]{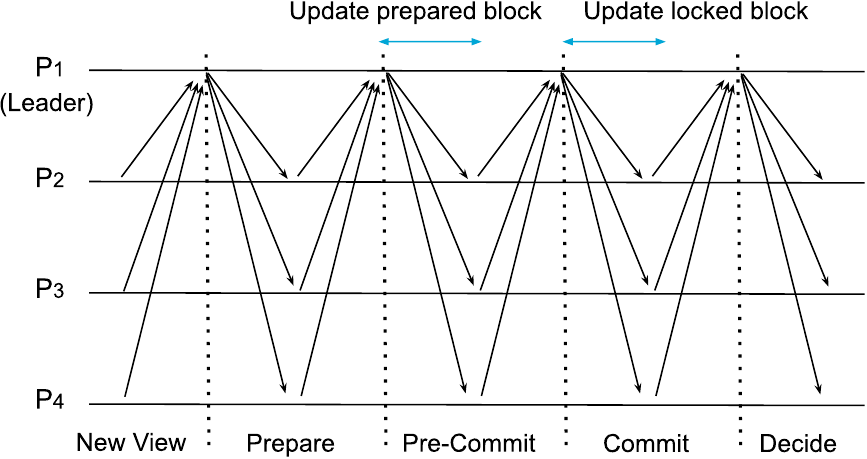}
    \caption{Communication phases of the HotStuff protocol.}
    \label{fig:hotstuff}
\end{figure}

HotStuff~\cite{hotstuff} is a leader-based BFT replication protocol whose message complexity is linear with the number of processes instead of quadratic as in PBFT~\cite{PBFT}. 
To achieve this goal, HotStuff's normal case consists in four communication phases that involve leader-to-replicas or replicas-to-leader communication, and its view-change procedure is embedded in its normal case. 
Classically, HotStuff uses $n=3f+1$ processes to tolerate $f$ faults and guarantees responsiveness because the leader initiates the next phase when it receives $n-f$ equal votes and because an unresponsive leader leads the replicas to initiate a view-change.  
HotStuff also mentions using threshold signatures and pipelining its operations to further improve its performance. Note that for simplicity, we focus on the non-pipelined version of HotStuff, which is called Basic HotStuff.  
The liveness and safety properties of HotStuff have been proven in partially synchronous networks, and testing our liveness checking tools on HotStuff allows us to evaluate possible false positives.   

HotStuff processes maintain and extend a chain of blocks that contain user transactions that are initialized with a Genesis block. All processes maintain the latest prepared and locked blocks they know of. In a nutshell, HotStuff proceeds according to five phases, which are illustrated in Figure~\ref{fig:hotstuff}. These phases can be described as follows.  

\noindent (1) \textit{New-view 1/2-phase.} All processes send the latest prepared block they know of to the leader. 

\noindent (2) \textit{Prepare phase.} The leader waits for $2f+1$ identical prepared blocks and sends a propose message to all processes that contain a block that extends it. All processes are expected to vote for this new block by sending to the leader their signature on it. In this phase, a process votes on a block if it extends the latest block it locked (for safety) or if it originates from a more recent view (for liveness). 

\noindent (3) \textit{Pre-commit phase.} The leader gathers $2f+1$ votes and aggregates them into a quorum certificate, which it sends to all processes. Upon receiving a quorum certificate, a process marks the block as being prepared and sends its vote for this block to the leader. 

\noindent (4) \textit{Commit phase.} The leader gathers a quorum certificate on a prepared block and forwards it to all processes. Upon receiving a quorum certificate in this phase, all processes mark this block as being locked and send their vote for it to the leader. 

\noindent (5) \textit{Decide 1/2-phase.} The leader assembles a quorum certificate (i.e., a set of $2f+1$ signatures on a block) on a locked block and forwards it to all processes. Upon receiving this quorum certificate all processes execute the block. 

\subsection{2-Phase HotStuff}

\begin{figure}[t]
  \centering
    \includegraphics[bb=0 0 390 229, width=.92\linewidth]{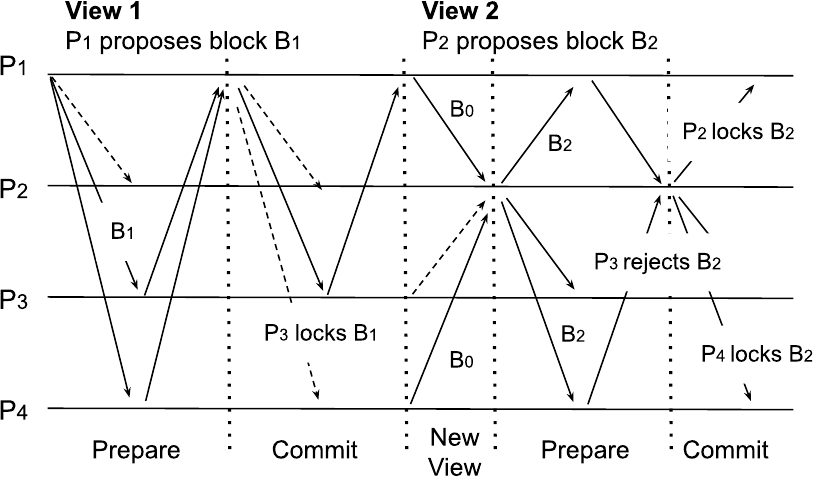}
    \caption{Example of a liveness violation in 2-phase HotStuff.}
    \label{fig:twophase_hotstuff}
\end{figure}

2-phase HotStuff is a variant of Basic HotStuff that Yin et al. discuss for pedagogical reasons in the original HotStuff paper~\cite{hotstuff}. 2-Phase HotStuff is very similar to  HotStuff and only differs from it by the fact that it combines the Precommit and Commit phases into a single phase. In 2-Phase HotStuff, a process can lock on a block once it receives a quorum certificate in the Prepare phase.  

However, this modification prevents 2-phase HotStuff from making progress in some particular scenarios, even though it remains safe. In these problematic scenarios, different processes lock on conflicting blocks and never get to execute any or update their locks. More precisely, in each view, a subset of processes locks on the block that is proposed while others reject it, and view-changes that are triggered by network asynchrony prevent sufficiently enough processes from adopting the newest proposed block. Under these circumstances, the system fails to progress and is stuck in an infinite loop. 

Figure~\ref{fig:twophase_hotstuff} illustrates one of these problematic scenarios with four processes $P_1$, $P_2$, $P_3$, and $P_4$. Process $P_1$ is faulty, while the other processes are correct. The scenario this Figure illustrates is the following. Originally, all processes are locked on the same block $B_0$ (e.g., the genesis block). 
In the first view, process $P_1$ is the leader and proposes a new block $B_1$ only to $P_3$ and $P_4$ in the Prepare phase, and then only sends the quorum certificate it assembles to $P_3$ in the Commit phase. The messages that $P_1$ omits to send are shown using dashes.  At the end of the first view, $P_3$ is the only correct process to lock on $B_1$. 
In the next view, $P_2$ is the leader. Processes $P_1$ and $P_4$ send block $B_0$ on which they locked to $P_2$, but $P_3$'s message, which contains block $B_1$, is delayed (shown using dashes). Consequently, $P_2$ proposes a block $B_2$ that extends over $B_0$ but is  in conflict with $B_1$ in the Prepare phase to all processes. Eventually, processes $P_2$ and $P_4$ lock on $B_2$ while $P_3$ rejects it and remains locked on $B_1$. 
The system is then deadlocked in future views if $P_1$ remains silent because no quorum of $2f+1$ processes can be assembled by any leader.  

\subsection{Sync HotStuff}

\begin{figure}
  \centering
  \begin{subfigure}[b]{\linewidth}
    \includegraphics[bb=0 0 382 212, width=.8\linewidth]{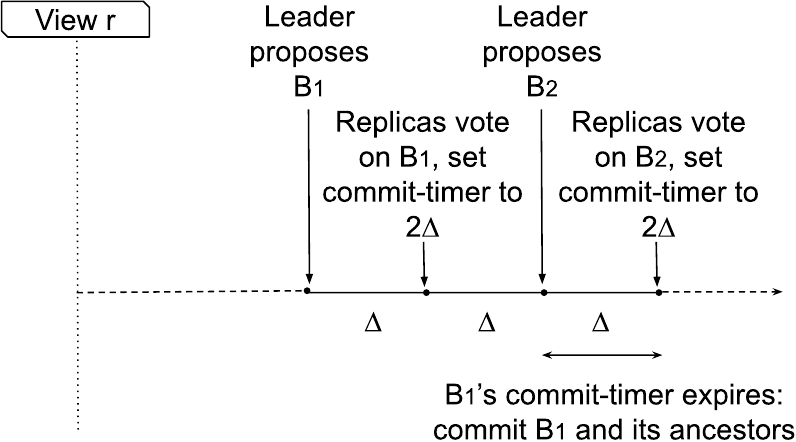}
    \caption{Steady case with 2 successive blocks $B_1$ and $B_2$. Block $B_1$ is committed by a process $2\Delta$ after it votes on it.}\label{fig:sync_normal}
  \end{subfigure}
  \vspace{5mm}
  \begin{subfigure}[b]{\linewidth}
    \includegraphics[bb=0 0 525 148, width=\linewidth]{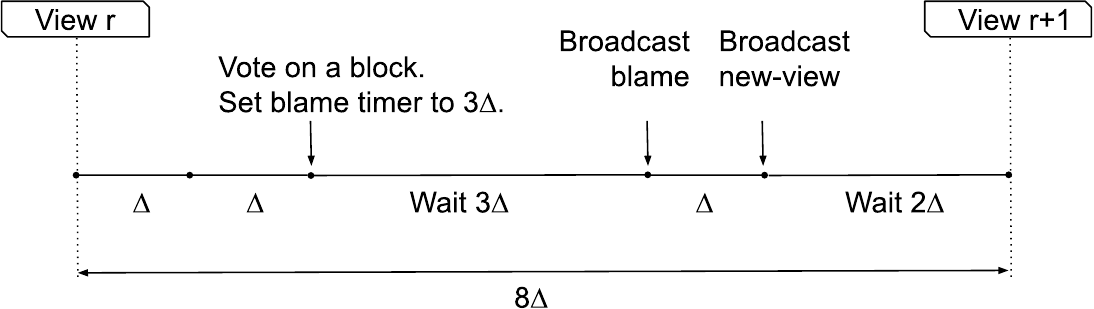}
    \caption{View change.}\label{fig:sync_vc}
  \end{subfigure}
  \caption{Sync HotStuff's steady case and view change procedures.}
\label{fig:synchotstuff}
\end{figure}

Sync HotStuff~\cite{sync_hotstuff} is a variant of HotStuff for synchronous networks. Sync HotStuff uses a minimum of $n=2f+1$ processes to tolerate $f$ Byzantine processes. Assuming that the communication latency is bounded by $\Delta$, Sync HotStuff's latency is bounded by $2\Delta$. Interestingly, Momose and Cruz's force-locking attack has shown that an adversary that controls the faulty processes and the network delays can break both the safety and the liveness of a preliminary version (eprint 20190312:115828) of Sync HotStuff~\cite{force_locking}. 
In this work, we discuss and focus on this early version of Sync HotStuff since it allows us to evaluate whether our liveness checking methods successfully detect its violations.  Figure~\ref{fig:synchotstuff} illustrates the steady case and the view change procedures of Sync HotStuff. 

In the steady case, upon entering a new view, all processes send their highest locked block to the new leader. After waiting for an initial $2\Delta$ period where it receives the highest locked blocks from all correct processes, an honest leader broadcasts a new block proposal that extends over the highest locked block along with a quorum certificate for the highest locked block to all processes. All honest processes subsequently broadcast their vote on the leader's proposal during the following round and initialize a local $2\Delta$ commit timer associated with this proposal. If a process does not detect a conflicting block when its commit timer expires, then it commits the block and all its ancestors and otherwise drops it. While waiting for commit timers to expire, the leader keeps proposing blocks, and processes keep voting on blocks, which is shown with blocks $B_1$ and $B_2$ in Figure~\ref{fig:sync_normal}. In the steady case, an honest leader, therefore, keeps proposing and committing blocks every $2\Delta$.

Whenever it votes on a block, a process resets a local blame timer to $3\Delta$. A view change (Figure~\ref{fig:sync_vc}) is triggered if a process's timer expires, if it refuses to vote on a block, or if detects that the leader broadcasts conflicting blocks, then it broadcasts a blame message. Upon receiving $f+1$ blames, a process broadcasts them and stops voting in view $r$, waits for $2\Delta$, and finally moves to view $r+1$.

Momose and Cruz's force-locking liveness and safety attacks exploit the fact that processes keep processing the messages they receive during the $2\Delta$ period that follows the reception of $f+1$ blames. 
During this period of time, some correct processes might receive a quorum certificate from a leader. In that case, these processes are then forced to update their locked blocks, while other honest processes might not be able to do so. If different honest processes are led to lock on conflicting blocks, then the system may never be able to make progress in the future, in particular, if faulty processes subsequently remain silent since no locked block can ever collect enough votes.

The safety attack on Sync HotStuff builds on the situation where honest processes have locked on different blocks. In subsequent views, the adversary is assumed to be able to leverage network delays and use the votes of the Byzantine processes to lead different subsets of correct processes to commit different blocks. 

\section{Liveness Checking Algorithms} 
\label{sec:method}

Checking safety properties during the execution of a protocol is straightforward since we can check whether each state that the system reaches violates them.  
Detecting liveness violations is more difficult than detecting safety properties since it requires finding an infinite execution that does not satisfy the system's properties. Existing testing approaches target the problem by either finding a sufficiently long execution or finding a lasso, i.e., an execution that can visit the same program state infinitely often.

In this section, we introduce two approaches that aim at identifying liveness violations. The first one is bounded liveness checking (also referred to as \emph{temperature checking})~\cite{killian2007life, partialcaching} and \emph{lasso detection with partial state caching}~\cite{partialcaching}. These methods build on the notion of a \emph{hot state}, which is a system-wide state in which the system does not satisfy some of its properties or produce useful results. Intuitively, the temperature-checking method reports a violation if the system remains in a hot state for a long time. On the other hand, lasso detection methods detect the existence of a cycle of states (i.e., a lasso) where the system can possibly get stuck forever following its transitions. 

\subsection{Temperature Checking}

Temperature checking checks for \emph{bounded} liveness, which means that it reports a violation if the execution does not produce a useful result (i.e., produce a new block in the case of blockchain consensus) for a specified amount of time. The method maintains a temperature variable $temp$, which is equal to the number of successive hot states the system remained in, and it reports a violation if the temperature reaches a certain threshold value \textit{TT} provided by the programmer. 

Algorithm~\ref{alg:checktemp} provides the pseudocode of temperature checking. The  $\checkTemp(s, trace, temp, TT)$ method is called whenever the system reaches a state $s$ after executing a sequence of system events $trace$. This function increments the system's temperature $temp$ if state $s$ is a hot state (line~\ref{line:incr}) and resets it to 0 otherwise (line~\ref{line:reset}). Violations are detected when the temperature reaches the temperature threshold (line~\ref{line:threshold}), and therefore directly depend on the value of parameter $TT$ that is provided by the user. A low threshold value might result in a high number of false positives, while a high threshold value leads to longer execution traces that are more difficult to interpret. 

\begin{algorithm}[t]
\setcounter{AlgoLine}{0}
\caption[CheckTemp]{Temperature checking.}
\label{alg:checktemp}

\DontPrintSemicolon
\SetKwProg{proc}{Procedure}{}{}

  \KwIn{Current state $s$}
  \KwIn{Current trace $trace$}
   \KwIn{Current temperature $temp$}
  \KwIn{Temperature threshold $TT$}
  \KwOut{Updated temperature value and trace}
  \BlankLine

\proc{$\checkTemp(s, trace, temp, TT)$}{ 
    \eIf{$\isHot(s)$}
    {
        $temp \gets temp + 1$\; \label{line:incr}
        
        \If{$temp = TT$} { \label{line:threshold}
            $\reportViol$($trace$)
        }
    }{
        $temp \gets 0$\; \label{line:reset}
    }
    \Return{$temp, trace$}

  \BlankLine
    }
\end{algorithm}

\subsection{Lasso Detection}

Liveness checking based on lasso-detection aims at finding a cycle of states (i.e., a lasso) where the system might get stuck and repeat its state infinitely often. 
 
Detecting lassos in the executions of distributed systems
is challenging because it is impractical to register the entire state of complex software systems. However, one can rely on the partial-state caching method~\cite{partialcaching}.
This method captures only part of the system state to check whether a partial state is repeated during an execution. Since the state caching is only partial, repeating the same partial state does not ensure repeating the same state in the execution. The existence of the cycle is then verified by replaying the execution of the detected trace using a controlled scheduler that enforces the execution of certain events and traces. 

Algorithm~\ref{alg:checkLasso_unfair} details the partial-state caching algorithm.  
Given an execution of a trace $trace$ that has reached a system state $s$, it checks whether the current execution may cause a liveness violation. To do so, it uses a hash function $Hash$ to hash the partial state information, which ideally maps each partial state to a different hash value.
For each new state $s$ that is reached during the execution, it then checks if $Hash(s)$ has been seen earlier (lines~\ref{line:forloop} and~\ref{line:hashtest}). If it is the case, then it means that a potential cycle in the state transition system has been identified. The cycle forms a liveness violation if the states in the traces do not satisfy the system's properties, i.e., if all the events $e \in trace$ are hot. We overload the method $\isHot$ that initially checks whether a state is hot to also check whether a trace is made of hot states (line~\ref{line:ishottrace}). If the algorithm finds a cycle with a hot trace, it then verifies the existence of a real cycle (line~\ref{line:cycle}) and, if so, reports a liveness violation. 
Note that differently from~\cite{partialcaching}, we do not check for fairness because we only require to observe the occurrence of a lasso to report a liveness violation and do not need to check for a fair cycle or for starvation.

\begin{algorithm}[t]
\setcounter{AlgoLine}{0}
\caption[checkLasso]{Lasso detection.}
\label{alg:checkLasso_unfair}

\DontPrintSemicolon
\SetKwProg{proc}{Procedure}{}{}

  \KwIn{Current state $s$}
  \KwIn{Current trace $Trace$}
  \KwOut{Updated trace}
  \BlankLine

 \proc{$\checkLasso(s, trace)$}{  
 
  \For{$i \leftarrow 0$ \KwTo len($trace$)} { \label{line:forloop}
    \If{Hash(s) = Hash($trace[i]$)} { \label{line:hashtest}
        $C \gets trace[i..len(trace)]$\;
        
        \If{$\isHot(C)$} { \label{line:ishottrace}
            \If{$\checkCycle(s, ..)$} { \label{line:cycle}
                $\reportViol$($trace$)
                }
            }
        }
    }
    \Return{$trace$} 
    
  \BlankLine 
  }
\end{algorithm}

\section{Liveness Checking of Streamlined Protocols} 
\label{sec:our-method}

While the temperature and partial state caching methods provide a practical solution for checking the liveness of software systems, they are not directly applicable to blockchain consensus systems for several reasons. 
First, there does not exist a common notion of a partial state that captures relevant information during the execution of blockchain systems. Defining partial states is a delicate task. On the one hand, a partial state that overly abstracts the system information may fail at capturing essential state information and therefore suffer from a high rate of false positives. On the other hand, a partial state that would include too much information would not be impractical with large software systems. 
Second, the notion of \emph{hot state} has not been defined for streamlined blockchain systems and is required by the temperature and lasso detection methods, which we aim to use.
Finally, the lasso detection method requires a controlled scheduler to check whether a detected potential cycle is replayable. More specifically, it uses the scheduler to enforce the system to run the sequence of events that produced the detected cycle of system states; it checks if the cycle is replayable and only reports a violation if it is replayable. This makes the lasso detection method difficult to apply for systems that do not have a controlled event scheduler.

In this work, we address these issues by: (i) formulating a \emph{partial-state} definition that captures the essential state information during the execution of a streamlined BFT consensus algorithm; (ii) defining \emph{hot states}, which are states that model bad states, for streamlined BFT consensus; and (iii) using the execution state space for checking the existence of lassos (i.e., state cycles).

\subsection{Partial State in the HotStuff Protocol Family}

Essentially, our partial process state encapsulates essential information about the state of a process, which is modified through the various phases of the protocol execution.
However, locked blocks are instrumental in known liveness violations. In addition, executed blocks allow us to identify situations where two locked blocks exist in the system, which is a necessary condition for liveness bugs, but one has been executed, which indicates progress. We, therefore, define partial process states as follows.

\vspace{2mm}
\begin{definition}[Partial process state] \label{def:hotstate}

We define the partial state $s$ of a process $p$ as a tuple $\tup{H(b_{prepared}), H(b_{lock}), H(b_{exec})}$ where $H(.)$ is a hash function, $b_{prepared}$ is the last block that the process prepared, $b_{lock}$ is the block that is locked by the process and $b_{exec}$ is the last block it executed. 

\end{definition}
\vspace{2mm}

As a check, since the liveness of the HotStuff protocols (cf. Section~\ref{sec:protocol}) can be violated in the presence of conflicting locks among the processes of a system, we verified that the protocol variables we included in partial states allow the detection of these liveness bugs. 
In the context of the HotStuff protocol, the system's state is defined as the set of states of all processes within the system. It is worth noting that, for the purpose of defining hot states, the state of the network channels may not be necessary to include.

\vspace{2mm}
\begin{definition}[Partial system state] \label{def:hotstate}
The partial system state is a vector $\tup{\textit{stateMap}}$, where $\textit{stateMap}: P \mapsto S$ maps each process $p$ to its partial state $S_p \in S$.
\end{definition}
\vspace{2mm}

Note that for performance reasons, we test whether two states are equal based on their hashes. The hash of a system state is computed once and stored along with it. 

\subsection{Hot State in the HotStuff Protocol Family}

We define hot states for streamlined consensus protocols 
based on their \emph{locking} phase, where a node locks on a block once it learns that a Byzantine quorum has committed to it. 

\vspace{2mm}
\begin{definition}[Hot state for streamlined protocols] \label{def:hotstate}
We say that a streamlined blockchain system is in a \emph{hot state} if it satisfies these conditions: (i) the correct processes hold at least two locks on conflicting blocks; (ii) there is no locked block on which a quorum certificate could be generated if all correct processes that have not locked on a block decided to lock on it; and (iii) a correct process has not executed a block. 
\end{definition}
\vspace{2mm}

Condition (ii) states that the processes are not able to generate a quorum certificate on one of the existing locked blocks to reach consensus in this view, i.e., no locked block can accumulate enough votes from processes that either already locked on it or could lock on it (because they have not locked a conflicting block). Condition (iii) guarantees that the system does not execute a block and responds to the clients.  

Given Definition~\ref{def:hotstate}, which provides a general definition of hot states for streamlined blockchain protocols, one can easily adapt the hot state definition for HotStuff, 2-Phase HotStuff, and Sync HotStuff given their quorum certificate sizes (i.e., $2f+1$ for HotStuff and 2-Phase HotStuff, and $f+1$ for Sync HotStuff). 

\paragraph*{Monitoring hot states} To check whether the system is in a hot state, we use a liveness monitor, which keeps track of the current state of the processes. In particular, we save the hashes of the \emph{prepared} $(b_{prepared}$), locked ($b_{lock}$), and the last executed block ($b_{exec}$) for each process, and we ignore other variables.  

For the temperature checking method, monitoring whether the system is in a hot state and maintaining the temperature variable to track the duration in which the system stays in a hot state is sufficient to detect potential liveness violations. For the lasso detection method, we additionally check whether the system can stay indefinitely in a hot state.

\subsection{Checking for Lassos}

Different from the standard approach for lasso detection, which detects cycles in the test executions and checks the repeatability of the cycle by trying to replay it, we use a different approach in this work. 
The main difficulty in replaying a cycle of system states is that it requires a controlled event scheduler, which can enforce the execution of a particular event to reach a certain system state in the execution. However, most distributed systems and testing frameworks do not have a controlled environment. For example, the Twins framework~\cite{twins}, which we use in this work, runs test execution scenarios with particular network faults and Byzantine behaviors, but it cannot control the execution of the protocol at the message granularity. Therefore, it cannot enforce the execution of a given schedule of events to check whether some detected cycle can be replayed. However, it is still possible to observe the states that the system reaches. 

In this work, we check for lassos on the \emph{state transition graph} of the system, which increases the likelihood of detecting potential cycles of states reachable in the executions and also does not require a controlled scheduler. We construct the state transition diagram using the information we collect in the test executions we run on the system.  
During each test execution, we collect the state reachability information about the observed (partial) states and build a state transition graph. 
Starting from the initial system state, we observe the system states that are reached after running \emph{a round of the protocol}. The execution of the protocol gives us a sequence of system states, where we transition from one state to another by running a protocol round. 

We maintain a single state transition graph $G = \tup{S, T}$ where $S$ keeps the set of observed system states and $T$ corresponds to the transitions between these states. 
The graph contains an edge from states $s_1$ to $s_2$ if we observed a transition from state $s_1$ to $s_2$ during one of the test executions. The state transition graph summarizes the set of states and the transitions between them that are encountered in a set of executions. We update the graph after each test execution with information about the new states and transitions. We made the choice of maintaining a single transition graph to increase the chances of observing a cycle. However, this choice implies that correctly identifying a cycle does not mean simply reaching a previously observed state since it does not necessarily create a cycle in the graph. Therefore, cycles have to be explicitly searched for in our state transition graph. It is sufficient to search for cycles after the state transition graph has been updated by all executions.

\begin{figure}[t]
  \centering
    \includegraphics[bb=0 0 328 249, width=0.85\linewidth]{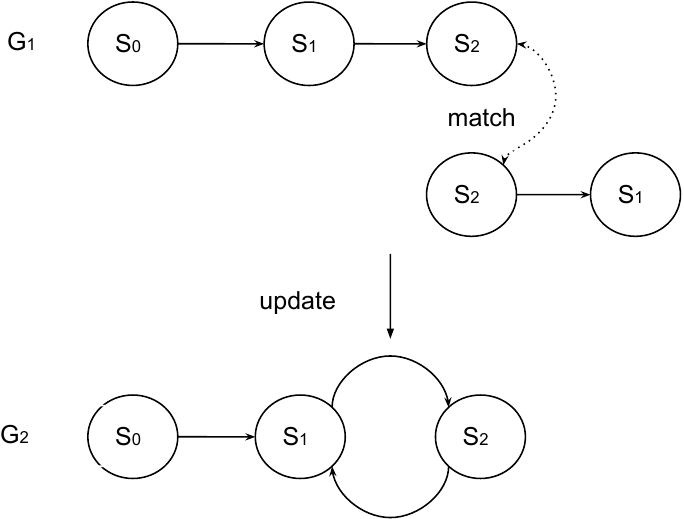}
    \caption{Updating the state transition graph after running a test execution.}
  \label{fig:update_cache}
\end{figure}

Figure~\ref{fig:update_cache} illustrates the state transition graph, its maintenance, and the appearance of a cycle in a simple example. In this example, a state transition graph $G_1$ keeps three system states $S_0$, $S_1$, and $S_2$, and their state transitions. Assuming that we run an additional test execution in which we observe that the system moves from $S_2$ to $S_1$, then we extend the graph by adding an edge from $S_2$ to $S_1$. The resulting graph $G_2$ contains a cycle between states $S_1$ and $S_2$, which is a potential lasso in the execution. 

\section{Implementation Details}
\label{sec:impl}

We now discuss our implementation of the checking methods and how we modified existing frameworks to test the liveness of HotStuff, Sync HotStuff, and 2-Phase HotStuff. 

\subsection{General Settings}

We built our experiments on top of the \texttt{relab/hotstuff} framework\footnote{\url{https://github.com/relab/hotstuff}}. This framework provides a useful set of auxiliary modules, such as digital signatures and encryption algorithms, and also provides some support for deploying distributed experiments via SSH. 

We generate the test executions using the Twins~\cite{twins} framework. 
We extended the Twins framework for our evaluation as follows.
First, we implemented the 2-phase HotStuff and Sync HotStuff protocols based on the consensus interfaces provided by the Twins framework. We rely on the implementation of HotStuff which is provided by Twins. Note that Fast HotStuff~\cite{fast_hotstuff} is also implemented in Twins. We do not consider Fast HotStuff in our experiments because it would be redundant with the use of HotStuff's: the liveness of both protocols has been demonstrated. 
Second, we implemented additional functions to capture the system state and maintain the state transition diagram. Our implementation of the state transition diagram is thread-safe to support the parallel execution of test scenarios.
Third, we extended the testing framework to generate and execute scenarios that can introduce message delays in the test executions. We used message-delaying test cases to test the executions of the Sync HotStuff protocol, whose correctness depends on the timing of the delivery of the messages.
Finally, we implemented the temperature and lasso detection methods for checking liveness.

\subsection{Implementation of Sync HotStuff} 

Among the protocols on which we apply our liveness-checking methods,  Sync HotStuff is the only one that cannot be directly simulated on top of Twins, since it does not rely on a lock-step process. More specifically, in the steady state, a leader in Sync HotStuff keeps proposing blocks every $2\Delta$ until any equivocation or network delays prevent progress. However, a Twins scenario requires a new leader to be specified for each view, while Sync HotStuff's steady operating mode uses a leader until a view-chance occurs. Therefore, we simulate Sync HotStuff on top of the view mechanisms that support the HotStuff protocols. 

\subsection{Extending Test Scenarios with Message Delays.}

The force-locking attack on Sync HotStuff is possible when messages can suffer from network delays (i.e., outside of the strictly synchronous network model). However, Twins does not support test scenarios with message delays and cannot detect these attacks. We extended Twins to generate test execution scenarios that delay the delivery of messages.
To trigger the force-locking attack, we inject message delays into the scenario generator. We randomly delay Propose messages with delays that belong in $[0, \frac{1}{2}\Delta, 1\Delta, \frac{3}{2}\Delta, 2\Delta, \frac{5}{2}\Delta, 3\Delta]$ and delay Vote messages with delays that belong in $[0, \frac{1}{2}\Delta, 1\Delta, \frac{3}{2}\Delta, 2\Delta]$. These values are chosen so that the delay of a Propose message does not exceed the value of starting a blame timer ($3\Delta$), otherwise, the proposal is dropped. The delay of a Vote message does not exceed the value of the commit timer ($2\Delta$), otherwise, the votes are too late, equivalent to Blame, and the leader can never collect a quorum of votes.

\section{Performance Evaluation}
\label{sec:eval}

We check the liveness of HotStuff, Sync HotStuff, and 2-Phase HotStuff by running a set of test executions and using the temperature and lasso detection methods to detect liveness violations in these executions. Then, we compare the performance of the temperature and lasso detection methods to a baseline bounded liveness checking method.

\subsection{Test Executions}

For the generation of test executions, we used the test scenarios generated by the Twins framework for the HotStuff family of protocols. 
We ran the test executions and checked their liveness using HP ZBook-Studio-G5 with a 2.6 GHz Intel Core i7 (12 cores), 16 GiB memory, and a UHD Graphics 630.

Table~\ref{tab:execution} provides the time required for Twins to execute a unit test that involves 10 or 20 blocks with HotStuff and Sync HotStuff. 
For each configuration, we randomly selected and executed 1.000 Twins unit test scenarios and report the average. All the scenarios are configured with 2 network partitions, 4 honest processes and 1 Byzantine (twin) process, and are executed for 10 or 20 rounds (blocks). For HotStuff, we apply a fixed network delay of 10\,ms. For the execution of Sync HotStuff, we set $\Delta$ to 50\,ms. 
The table lists the average execution times $T_{mean}$ and the standard deviations $T_{std}$ for each protocol and number of blocks.
The times required to execute a scenario with Sync HotStuff are larger than those of 2-Phase HotStuff and HotStuff since Sync HotStuff requires the use of longer network delays and malicious delaying by the adversary. For example, with 20 blocks, Sync HotStuff requires 4.156\,s to execute a scenario, while 2-Phase HotStuff and HotStuff respectively require 2.194\,s (47\% less) and 2.813\,s (32\% less).  

\begin{table}[t]
\normalsize
\centering
\renewcommand{\arraystretch}{1.1}
\begin{tabular}{ |c|c|c|c| } 
\hline
Protocol & Rounds & $T_{mean}$ (s) & $T_{std}$ (s) \\
\hline
\multirow{2}{8em}{2-Phase HotStuff} 
& 10 & 0.759 & 0.351 \\ 
& 20 & 2.194 & 1.398 \\ 
\hline
\multirow{2}{8em}{HotStuff} 
& 10 & 0.838 & 0.245 \\ 
& 20 & 2.813 & 0.919 \\ 
\hline
\multirow{2}{8em}{Sync HotStuff} 
& 10 & 3.100 & 1.062 \\ 
& 20 & 4.156 & 2.151 \\ 
\hline
\end{tabular}
\vspace{1mm}
\caption{Execution time of a unit test scenario in seconds under various configurations.}
\label{tab:execution}
\end{table}

\subsection{Time-Bounded Liveness Checking} 

As a baseline, we checked the bounded liveness of the consensus algorithm executions using a time bound. This method takes a time bound parameter from the programmer and reports a potential liveness violation if the system does not reach a consensus within the given specified delay.  

The effectiveness of the time-bounded liveness checking method depends on the actual value of the time-bound parameter chosen by the programmer. In our evaluation, we selected three representative bound values per protocol based on its expected normal case execution time, which we have presented in Table~\ref{tab:execution}.  
We utilized different values for the time-bound parameter: 
i) a small bound value $T_{small} = T_{mean}$; 
ii) an intermediate bound value  $T_{mid} = T_{mean} + T_{std}$, which covers 84\% of the values of a normal distribution; 
iii) a large bound value $T_{large} = T_{mean} + 2T_{std}$, which covers 98\% of the values of a normal distribution. 

For a given time-bound parameter value, an execution is expected to reach consensus before the time-bound and, if not, will be associated with a potential liveness violation.  
Consequently, one can expect that increasing the value of the time-bound parameter decreases the number of false positive liveness violations, but it also increases the computational overhead. Selecting the right time-bound value is, therefore, a delicate process.  

\subsection{False positives}

False positive liveness violations can be detected based on the analysis of the replicas' local variables, such as their view number, prepared blocks, and locked blocks. However, the analysis of the remaining positives requires a thorough manual evaluation of the execution by developers. Indeed, this analysis is not straightforward, because network partitions and message drops might prevent the processes' view numbers to not be synchronized. 
We observe that the identification of false positives consists in distinguishing whether there are really two conflicting chains in the system so that it can no longer make progress. 

In our evaluation, we consider an execution to be a false positive if it does not disallow progress, that is, if the execution does not keep two conflicting locked blocks in the distributed processes, or if there are two locked blocks such that one is extending the other. 

\begin{table*}[t]
\normalsize
\centering
\renewcommand{\arraystretch}{1}
\begin{adjustbox}{max width=\textwidth}
    \begin{tabular}{ *{8}{|c}| } 
    \hline
     \thead{Rounds} & \thead{Method} & \thead{Threshold} & \thead{Time} & \thead{Trace length} & \thead{\% Safety \\ violations} & \thead{\% Liveness \\ violations} & \thead{\% False \\ positives} \\
    \hline
    \hline
    \multirow{4}*{10} 
    & Temperature & 5 & 17 min 3 s  & - & 0 & 0 & 0 \\ 
    \cline{2-8}
    & Lasso detection & - & 17 min 16 s & - & 0 & 0 & 0 \\ 
    \cline{2-8}
    & Small-Timeout & 0.8s & 14 min 50 s  & 8 & 0 & 98.8 & 100\\ 
    \cline{2-8}
    & Mid-Timeout & 1.2s & 20 min 23 s & 9 & 0 & 96.4 & 100 \\ 
    \cline{2-8}
    & Large-Timeout & 1.6s & 31 min 43 s & 10 & 0 & 94.3 & 100 \\ 
    \hline
    \hline
    \multirow{6}*{20}
    & Temperature & 5 & 52 min 26 s & - & 0 & 0 & 0 \\ 
    \cline{2-8}
    & Lasso detection & - & 57 min 3 s & - & 0 & 0 & 0 \\ 
    \cline{2-8}
    & Small-Timeout & 2.8s & 48 min 43 s & 15 & 0 & 95.7 & 100 \\ 
    \cline{2-8}
    & Mid-Timeout & 3.8s & 1 h 5 min 12 s & 18 & 0 & 94.1 & 100 \\ 
    \cline{2-8}
    & Large-Timeout & 4.8s & 1 h 19 min 34 s & 20 & 0 & 90.6 & 100 \\ 
    \hline
    \end{tabular}
\end{adjustbox}
\vspace{1mm}
\caption{Liveness and safety violations detected with the temperature checking, lasso detection, and bounded liveness methods on executions of the HotStuff protocol.}
\label{tab:hotstuff}
\end{table*}
\begin{table*}[t]
\normalsize
\centering
\renewcommand{\arraystretch}{1}
\begin{adjustbox}{max width=\textwidth}
    \begin{tabular}{ *{8}{|c}| } 
    \hline
     \thead{Rounds} & \thead{Method} & \thead{Threshold} & \thead{Time} & \thead{Trace length} & \thead{\% Safety \\ violations} & \thead{\% Liveness \\ violations} & \thead{\% False \\ positives} \\
    \hline
    \hline
    \multirow{4}*{10} & Temperature & 5 & 16 min 25 s & 9 & 0 & 0.23 & 0\\ 
    \cline{2-8}
    & Lasso detection & - & 16 min 19 s & 8 & 0 & 0.42 & 0\\ 
    \cline{2-8}
    & Small-Timeout & 0.8s & 14 min 21 s & 8 & 0 & 77.8 & 98.2\\ 
    \cline{2-8}
    & Mid-Timeout & 1.2s & 19 min 44 s & 9 & 0 & 74.6 & 94.5\\ 
    \cline{2-8}
    & Large-Timeout & 1.6s & 30 min 11 s & 10 & 0 & 58.6 & 88.6\\ 
    \hline
    \hline
    \multirow{6}*{20} 
    & \multirow{3}*{Temperature} & 5 & 51 min 13 s & 12 & 0 & 1.92 & 0\\ 
    \cline{3-8}
    &  & 10 & 53 min 30 s  & 17 & 0 & 0.74 & 0\\ 
    \cline{3-8}
    &  & 15 & 54 min 2 s & 20 & 0 & 0.17 & 0\\ 
    \cline{2-8}
    & Lasso detection & - & 52 min 26 s & 13 & 0 & 2.04 & 0\\ 
    \cline{2-8}
    & Small-Timeout & 2.2s & 38 min 52 s & 12 & 0 & 78.8 & 97.6\\ 
    \cline{2-8}
    & Mid-Timeout & 3.6s & 59 min 12 s & 16 & 0 & 66.4 & 90.4\\ 
    \cline{2-8}
    & Large-Timeout & 5.0s & 1h 20 min 33 s & 20 & 0 & 46.5 & 77.9\\ 
    \hline
    \end{tabular}
\end{adjustbox}
\vspace{1mm}
\caption{Liveness and safety violations detected with the temperature checking, lasso detection, and bounded liveness methods on executions of the 2-Phase HotStuff protocol.}
\label{tab:twophaseHotStuff}
\end{table*}
\begin{table*}[t]
\normalsize
\centering
\renewcommand{\arraystretch}{1}
\begin{adjustbox}{max width=\textwidth}
    \begin{tabular}{ *{8}{|c}| } 
    \hline
     \thead{Rounds} & \thead{Method} & \thead{Threshold} & \thead{Time} & \thead{Trace length} & \thead{\% Safety \\ violations} & \thead{\% Liveness \\ violations} & \thead{\% False \\ positives} \\
    \hline
    \hline
    \multirow{4}*{10} & Temperature & 5 & 6 min 54 s  & 7 & 3.1 & 1.8 & 2.2\\ 
    \cline{2-8}
    & Lasso detection & - & 6 min 38 s & 6 & 2.6 & 1.3 & 2.6\\ 
    \cline{2-8}
    & Small-Timeout & 3.1s & 5 min 49 s & 7 & 0.3 & 33.6 & 96.4\\ 
    \cline{2-8}
    & Mid-Timeout & 4.1s & 6 min 41 s & 8 & 2.5 & 32.4 & 92.7\\ 
    \cline{2-8}
    & Large-Timeout & 5.1s & 8 min 37 s & 10 & 2.6 & 20.7 & 84.8\\ 
    \hline
    \hline
    \multirow{6}*{20} & \multirow{3}*{Temperature} & 5 & 8 min 36 s & 9 & 1.7 & 1.9 & 3.6 \\ 
    \cline{3-8}
    &  & 10 & 9 min 4 s & 14 & 1.1 & 0.5 & 0\\ 
    \cline{3-8}
    &  & 15 & 9 min 12 s & 18 & 0.7 & 0.1 & 0\\ 
    \cline{2-8}
    & Lasso detection & - & 8 min 42 s & 8 & 2.1 & 1.6 & 3.4\\ 
    \cline{2-8}
    & Small-Timeout & 4.2s & 7 min 2 s & 7 & 0 & 34.2 & 95.3\\ 
    \cline{2-8}
    & Mid-Timeout & 6.4s & 10 min 24 s & 15 & 0.3 & 24.2 & 91.2\\ 
    \cline{2-8}
    & Large-Timeout & 8.6s & 14 min 9 s & 20 & 1.6 & 16.7 & 81.2\\ 
    \hline
    \end{tabular}
\end{adjustbox}
\vspace{1mm}
\caption{Liveness and safety violations detected with the temperature checking, lasso detection, and bounded liveness methods on executions of the Sync HotStuff protocol.}
\label{tab:synchotstuff_50ms}
\end{table*}

\subsection{Checking HotStuff's Liveness}

Table~\ref{tab:hotstuff} lists the results of testing the HotStuff protocol with 10,000 randomly selected Twins test scenarios for 10 and 20 rounds and checking the liveness of the executions using the temperature and lasso-detection based methods, along with the baseline time-bound checking.
The columns show the value of the threshold parameter (temperature threshold for the temperature method and time threshold for the time-bounded checking method), the total runtime to run the tests (in seconds), the average number of rounds to report a violation (or a dash "-" if there are no violations), and the ratios of safety and liveness violations detected in the executions. For the safety property, we checked the agreement of the processes by comparing their executed blocks. For the liveness property, we used time-bound and temperature methods to check whether consensus is reached within a bounded duration of execution (bounded by temperature and time, respectively) and the lasso detection method to check whether the system can stay in a cycle of hot system states and therefore does not satisfy its property.
 
For the temperature method, we estimated a temperature threshold equal to 5 rounds, i.e., we report a violation if the system states in a hot state for 5 rounds. We use the same temperature threshold for 2-Phase HotStuff and Sync HotStuff protocols.   
With this temperature threshold value, the temperature-based and the lasso detection-based methods did not report any violations for HotStuff (i.e., 0 in the safety and liveness violations columns). 

On the other hand, the time-bounded liveness checking baseline reported many potential liveness violations, where the executions could not reach a consensus in the given amount of time. 
The results for time-bounded liveness checking show that using a small timeout reports a lot of violations. For example, with 20 rounds and the small timeout value, the time-bounded liveness checking method reported that 78.8\% of the executions contain liveness violations. 
However, 
97.6\% of those are false positives where consensus has not been reached within the allocated number of rounds because of network partitions. 
In fact, it is very likely that the Twins test generator produces scenarios that cannot gather a quorum of votes because of network partitions and lack of leader replacement. Most of such test scenarios have not even completed their executions before timing out, and then they are labeled as violations. 
Albeit fewer, using a larger timeout bound still leads to reporting a high number of false positive liveness violations in which the system does not even enter a hot state. For example, still with 20 rounds, using the large timeout value decreases the proportion of executions that contain liveness violations to 90.6\%. 
Overall, time-bounded checking reports them as potential liveness violations since these scenarios cannot reach a consensus due to frequent network partitions. 

\subsection{Checking 2-Phase HotStuff's Liveness}

Table~\ref{tab:twophaseHotStuff} lists the results we obtained by executing randomly selected unit test scenarios on 2-Phase HotStuff, which is known to violate liveness in certain scenarios. 

Similar to the results for HotStuff, we observe that time-bounded liveness checking methods provide a higher amount of false positives than temperature and lasso-based methods. 
To evaluate the effect of the temperature value on the amount of reported false positives, we ran the tests for a varying number of temperature bounds (5, 10, 15) and checked if the system execution can reach those temperature bounds in the execution of 20 rounds.
We observed that increasing the temperature threshold increases the required time to complete the test executions while it reduces the amount of reported false positives. For example, replacing the small timeout by the large timeout with 20 rounds increased the computation time from 38\,min 52\,s to 1\,h 20\,min 33\,s, and reduced the proportion of false positives from 97.6\% to 77.9\%. 
Similarly, the time taken for the time-bounded checking also increases with the time-bound. 
For the lasso detection method, the execution time does not depend on a predefined bound but on the execution time of the test case together with the time required for cycle detection. 

\subsection{Checking Sync HotStuff's Liveness}

Table~\ref{tab:synchotstuff_50ms} shows the execution results of testing Sync HotStuff (eprint version 20190312:115828) with $\Delta = 50\,\mathrm{ms}$. 
Our findings are mostly identical to the observations we made with HotStuff (Table~\ref{tab:hotstuff}) and 2-Phase HotStuff (Table~\ref{tab:twophaseHotStuff}). 
One significant difference, however, is the fact that this version of Sync HotStuff can violate both safety and liveness in certain test scenarios. For example, we found that the lasso detection method identified that 2.1\% and 1.6\% of the executions respectively contained safety and liveness violations with 20 rounds. 
Additionally, we have observed slightly more false positive liveness violations in Sync HotStuff than with HotStuff and 2-Phase HotStuff. We believe that this higher false positive rate comes from the adaptations we had to make to run Sync HotStuff on top of Twins.  

\subsection{Discussion}

The resource consumption of all methods are reasonable. For example, all experiments used less than 65 MB of memory, and consumed around 12\% of a 2.6 GHz Intel Core i7 CPU. 

More importantly, time-bounded liveness checking introduces a significant number of false positives when applied to streamlined protocols, as compared to temperature-based and lasso-based techniques. We find that using a small timeout for liveness checking is not practical, as it fails to detect actual violations; rather, it detects a considerable number of false positives. 
Even when increasing the timeout duration to a larger value, the rate of false positives remains high, and the execution time significantly increases. In contrast, temperature-based and lasso-based techniques utilize the concept of hot states to detect the occurrence of a violation, making them more precise and reducing the likelihood of false positives. 

\section{Conclusion}
\label{sec:conc}

We have investigated the existing liveness violations in streamlined Byzantine consensus protocols, and more specifically, in the HotStuff protocol family. We found out that existing time bounded checking methods generate a high number of false positives. To overcome this limitation, we have adopted temperature and lasso-detection liveness-checking techniques to streamlined Byzantine consensus protocols, for the first time. We focused on the HotStuff protocol family and propose necessary definitions of system state abstractions (namely partial and hot states). Our approach can be generalized to other BFT consensus protocols.  Furthermore, we demonstrate that the implementation of our methods on top of the Twins testing framework successfully detects the liveness violations in 2-Phase HotStuff and in an early version of Sync HotStuff. 
Both the temperature and the lasso-detection methods are shown to be practical for the liveness-checking of blockchain consensus algorithms. Our results indicate that these methods identify liveness violations with fewer false positives than the time-bounded liveness checking baseline.
However, 
adequately selecting the threshold of the temperature-based method is a delicate task that can lead to false positives or high computational overhead, while lasso detection might be more memory intensive for complex systems.

\printbibliography

\end{document}